\documentclass[english]{article}
\usepackage[T1]{fontenc}
\usepackage[latin9]{inputenc}
\usepackage{array}
\usepackage{float}
\usepackage{multirow}
\usepackage{amstext}
\usepackage{graphicx}
\usepackage[authoryear]{natbib}

\makeatletter

\providecommand{\tabularnewline}{\\}

\newcommand{\lyxaddress}[1]{
	\par {\raggedright #1
	\vspace{1.4em}
	\noindent\par}
}

\makeatother

\usepackage{babel}
\begin{document}
\title{Cyberattacks on Quantum Networked Computation and Communications --
Hacking the Superdense Coding Protocol on IBM\textquoteright s Quantum
Computers}
\author{Carlos Pedro Gonçalves}
\maketitle

\lyxaddress{Lusophone University of Humanities and Technologies, e-mail: p6186@ulusofona.pt}
\begin{abstract}
The development of automated gate specification for quantum communications
and quantum networked computation opens up the way for malware designed
at corrupting the automation software, changing the automated quantum
communications protocols and algorithms. We study two types of attacks
on automated quantum communications protocols and simulate these attacks
on the superdense coding protocol, using remote access to IBM\textquoteright s
Quantum Computers available through IBM Q Experience to simulate these
attacks on what would be a low noise quantum communications network.
The first type of attack leads to a hacker-controlled bijective transformation
of the final measured strings, the second type of attack is a unitary
scrambling attack that modifies the automated gate specification to
effectively scramble the final measurement, disrupting quantum communications
and taking advantage of quantum randomness upon measurement in a way
that makes it difficult to distinguish from hardware malfunction or
from a sudden rise in environmental noise. We show that, due to quantum
entanglement and symmetries, the second type of attack works as a
way to strategically disrupt quantum communications networks and quantum
networked computation in a way that makes it difficult to ascertain
which node was attacked. The main findings are discussed in the wider
setting of quantum cybersecurity and quantum networked computation,
where ways of hacking including the role of insider threats are discussed.
\end{abstract}
\textbf{Ketwords:} Quantum Networked Computation, Quantum Internet,
Quantum Cybersecurity, Entanglement, Symmetry, Automated Quantum Communications,
Superdense Coding, IBM Q Experience, Intelligence Studies, Strategic
Studies

\section{Introduction}

The current interface with quantum computing resources mainly relies
on a cloud-based access to quantum computing, one can design quantum
algorithms and see the drawing of the circuits on a computer screen
or look at some implementation of this code in some programming language,
a good example of this type of interface is IBM\textquoteright s Quantum
Experience (IBM Q Experience), which uses a visual circuit design
operating with Open Quantum Assembly Language (OpenQASM) and Qiskit,
which allows one to use the Python programming language for implementing
quantum algorithms that can be run through cloud-based access to quantum
computers, including the simulation of quantum networked computation,
quantum communications and, even, quantum machine learning (Cross
et al., 2017; Cross, 2018; Gonçalves, 2019).

With the development of quantum computation and communications, cloud-based
access to quantum servers and an access to quantum communications
and computational networks, combining elements of a classical internet
and a quantum internet, may become increasingly feasible (Gonçalves,
2017, 2019), furthermore, the development of quantum communications
and remote access to quantum computation will demand some level of
automation of communications\textquoteright{} protocols and of computation
networks, where a classical level user interaction with quantum technologies
will tend to involve a standard high-level user interface, under which
that user\textquoteright s interaction and commands are automatically
encoded into quantum machine language, without the user having to
think about the machine language implementation or even know quantum
computation.

Such an infrastructure demands an automated translation of a user\textquoteright s
interaction patterns and intended actions into qubits and relevant
quantum circuits, as well as the automation of quantum communications
protocols and algorithms, which raises the possibility of malware
that can take advantage and disrupt the automated translation of high-level
interactions into quantum machine language, effectively disrupting
the interaction with quantum devices and quantum communications. This
software-level hacking, which draws upon the fact that such communications\textquoteright{}
protocols involve automated quantum networked computation that demands
appropriate software tools to automatically implement the correct
computations at each network node for the communications protocols
to function, is what we consider here, within the context of quantum
cybersecurity.

The research field of quantum cybersecurity, within the context of
Strategic Studies, includes the assessment of the strategic advantage
coming from quantum networked computation in terms of cyberdefense,
cryptographic solutions and information dominance (Mailloux et al.,
2016a,b; Gonçalves, 2017, 2019; Abellan and Pruneri, 2018; Gompert
and Libicki, 2021), it also addresses cybersecurity threats coming
from quantum computers as well as the cybersecurity threats to quantum
networks including attacks on quantum repeaters and quantum key distribution
(QKD) (Wu and Lidar, 2006; Larsson, 2002; Lydersen et al., 2010; Gerhardt
et al., 2011; Jogenfors et al., 2015; Mailloux et al., 2016b, Hughes-Salas
et al., 2018; Makarov and Hjelme, 2005; Satoh et al., 2018, 2020).

Indeed, QKD has been shown to be vulnerable to Distributed Denial
of Service (DDoS) attacks (Schartner and Rass, 2010) that disrupt
key generation, proposals to reduce this vulnerability include quantum-secured
paths over a network configuration, Hughes-Salas et al. (2018), for
instance, proposed a DDoS mitigation over a QKD network using software
defined networking (SDN), testing experimentally not only the vulnerability
of QKD to DDoS attacks but also the role of an SDN application for
mitigating these attacks.

While eavesdropping on quantum communications is an important goal
for hackers, other types of cyberattacks on quantum computers and
communications are possible for which a hacker\textquoteright s intention
is not to eavesdrop but rather to disrupt quantum networked computation
and quantum communications (Wu and Lidar, 2006).

In a setting where quantum communications protocols can be automated,
using software for automatically translating high-level interface
interactions into sequences of unitary gates to be implemented at
different nodes of a quantum communications network, the possibility
of an attack to disrupt common quantum communications protocols becomes
feasible, such that the software that was used to automate a given
network node\textquoteright s quantum computations on received qubits
can be changed by malware installed either remotely or by way of an
insider, such a malware can change the software code so that the main
automated communications protocol no longer runs the protocol\textquoteright s
quantum networked computational circuit but, instead, runs a different
(hacked) circuit, in this way, the automated communications protocol
is disrupted, it is this scenario that we deal with in the present
work.

These types of attacks fall in the same typology of DDoS in terms
of pattern, in the sense that their strategic goal is to disrupt networked
computation and communications during a period of time, for a strategic
or tactical advantage, these types of attacks are not aimed at eavesdropping,
nor, like DDoS, are they supposed to be hidden, in the sense that
the attacked parties know or may quickly find out that they are being
attacked, but the problem of finding which network node was attacked
becomes more difficult due to the use of entanglement and symmetries
in quantum communications, which allows hackers to hide the attacked
nodes and qubits, a point that we will address in the studied examples.

Like a classical computer virus that can corrupt a system\textquoteright s
function, malware targeting automation of quantum communications and
networked computation protocols may become the next frontier in hacking
when faced with a sufficiently advanced quantum computational and
communicational infrastructure, where automation for translation of
high-level commands to quantum machine language is hacked.

Under this context, if a quantum communications protocol has been
hacked with installed malware this may lead to a corruption of the
interaction with the networked quantum computational and communicational
infrastructure. From a hacker\textquoteright s standpoint, this is
not about eavesdropping on a quantum communications channel, as stated
above, but rather about installing malware that can disrupt quantum
communications. In the current article, we deal with this framework,
such that the hacking is assumed to be done by replacement of code
for automated gate specification in automated communications protocols.
We analyze different attack patterns to a well-known quantum communications
protocol: quantum superdense coding. 

We assume, as stated, an automated setting where the users do not
directly access the quantum gates but, instead, operate on a high-level
classical user interface, such that the whole translation of the classical
message to the quantum framework is done automatically. In this setting,
a hacker can attack different nodes in the communications network
in order to change the sent message without eavesdropping on the circuit,
leading to a bijective recoding of the message to be sent, this is
the first type of attack to the superdense coding protocol that we
address here, we call it a bijection attack.

Bijection attacks can take longer to detect, unless there is communication
between Alice and Bob with test qubits sent to evaluate whether or
not the communication circuit has been hacked or if there is a disruption
coming from post-processing automation that relies on the protocol\textquoteright s
output qubits. Bijection attacks are easier to deal with, in that
one can effectively invert the bijection to recover the original message,
quarantining the malware and giving enough time for a quantum cybersecurity
team to find the malware and remove it.

A second type of attack that we research is the purposeful scrambling
of the communications network, introducing specific unitary gates
that lead to a random output for the circuit at the measurement endpoint.
This second type of attack, which also uses unitary gates, randomizes
the final decoded message, taking advantage of quantum randomness
upon measurement, the main goal of the hacker is, again, not to eavesdrop
but rather to scramble the decoded message at the end of the communications
protocol when measurement takes place, this attack is not meant to
be hidden, it becomes a quantum variant of a DDoS attack, in the sense
that it makes the network\textquoteright s quantum computational and
communicational operativity useless, being difficult to distinguish
from a hardware malfunction or an environmental increase in noise.

We also show that, due to entanglement and symmetries, the hacker
can produce random results, upon measurement, on a final communications\textquoteright{}
node where qubits are measured, by attacking another node and changing
other qubits in a communications network. In a context of automated
communications protocols, where users have a high-level classical
interface with the communications infrastructure, it is up to quantum
cybersecurity to find the nodes and qubits that were hacked, but entanglement
and symmetries become a problem in the sense that the same end-result
can occur by attacking different nodes and qubits with different quantum
unitary gates, which demands, on the part of quantum cybersecurity,
a checking of different elements in a quantum communications network.

As stated, this attack cannot be easily quarantined in the way the
previous one was, especially since different unitary gates applied
to different nodes and qubits can, as we will show, lead to the same
end result, for instance, the hacker can attack one node and one qubit
with the noise-like results occurring at another node and for a different
qubit, this places an added pressure on quantum cybersecurity forensics
to find the attacked networked node and remove the malware. While,
in the case of the simpler quantum superdense coding protocol, this
checking is easier to do, since we have three nodes that can be hacked
(the entanglement source, Alice\textquoteright s system and Bob\textquoteright s
system), for larger communications networks, the number of nodes that
need to be checked for malware rises. Scrambling attacks can, therefore,
become just as damaging as a DDoS attack.

The work is divided into two sections, in section 2, we review the
superdense coding protocol (subsection 2.1), then we address the bijection
attacks (subsection 2.2) and the scrambling attacks (subsection 2.3),
simulating these attacks on IBM\textquoteright s quantum computers,
in section 3, we present a final discussion on the implications for
quantum cybersecurity and the demand for the development of quantum
cybersecurity forensics as a research field that can study how quantum
cyberattacks aimed at disrupting quantum communications and quantum
computation can be implemented, as well as how to find methods for
detection, protection and elimination of quantum malware. 

On this regard, there is an intersection with Intelligence Studies,
in that a main way for implementing the types of attacks that we are
considering, at least in the foreseeable future, comes from using
human intelligence (HUMINT) in the form of an infiltrated agent, or
compromised employee in a target organization, in order to find vulnerabilities
and install the malware to disrupt the automated protocols, in this
way, stronger cybersecurity countermeasures linked to insider threat
are needed in order to protect quantum networked computation from
disruption. This is a critical subject matter for countries\textquoteright{}
National Intelligence and Security since, as distributed quantum computation
and a hybrid of quantum and classical internet is developed, Universities,
Corporations, Banks, Governments, Armed Forces and Intelligence Agencies
will be at the forefront as targets for maneuvers based on HUMINT
aimed at disrupting quantum communications protocols.

\section{Hacking Superdense Coding}

\subsection{Superdense Coding}

The superdense coding protocol uses entanglement, symmetry and quantum
interference to allow Alice to communicate a two classical bits message
to Bob, sending only one qubit (Zygelman, 2018). In the framework
that we are considering, there is an automated source that entangles
two qubits and sends one to Alice\textquoteright s automated system
and another one to Bob\textquoteright s automated system, so that
Alice and Bob\textquoteright s respective systems share a symmetric
Bell pair of the type:

\begin{equation}
\left|\textrm{Bob+Alice}\right\rangle =\frac{\left|00\right\rangle +\left|11\right\rangle }{\sqrt{2}}
\end{equation}

Then Alice\textquoteright s system, as per the protocol, performs
automatically a series of computations on the qubit it received from
the entanglement source, depending on the message to be sent, these
computations follow the automated superdense coding protocol, after
these computations, Alice\textquoteright s system sends the transformed
qubit to Bob\textquoteright s system which automatically applies a
CNOT gate on the qubit it received from the entanglement source using
the qubit it received from the Alice\textquoteright s system as the
control qubit, after this CNOT gate, Bob\textquoteright s system performs
a Hadamard transform on Alice\textquoteright s qubit and then measures
both qubits to extract the final bit string, which is decoded and
shown to Bob as a high-level message.

As discussed in the introduction, we consider here a context of quantum
communications where the users operate on a high-level interface,
with the quantum gates being automatically implemented in the background
by software without the users accessing them, that is, we are considering
a framework where quantum computation and quantum communications have
become sufficiently developed so that there is an automation of the
translation of high-level instructions to the (quantum) machine language,
so that Alice just types in the message, in the case of superdense
coding one of four possible messages that she wants to send Bob, which
is then translated into one of four corresponding sequences of quantum
operations that will lead to the desired result under the established
automated superdense coding communications protocol.

In the context of superdense coding, Alice can send one of four alternative
messages to Bob, encoded in the machine language by a two bits string
or, under repeated use of the quatum communications network, longer
messages represented by longer strings. To keep things general, we
are not addressing the specific messages that Alice may be sending,
therefore, we just assume that Alice sends one of four messages, and
that the system automatically differentiates between one of those
four messages using the classical bit strings in the set \{00,01,10,11\},
the translation on Bob's end from these bit strings to actual high-level
messages is not also being addressed here, we will however discuss
the issue of the uncovering of the hack, if that hack occurs, which
will demand, on the part of a quatum cybersecurity forensics team,
to look at the systems\textquoteright{} automation software for installed
malware.

Considering the standard superdense coding without the hack, in the
general scenario under analysis, the adaptive quantum circuit composition
is fully automated and is like a black box for both Alice and Bob,
who are, in this scenario, like standard users that know nothing about
the actual workings of quantum computation nor do they care about
it, all they know is how to interact with the high-level interface
not looking \textquotedbl under the hood\textquotedbl .

The above is a very important point, we are assuming an advanced stage
of automation and integration of cyber-physical systems with quantum
communications' infrastructures, in such a way that the people in
the communication circuit are not quantum experts and the whole quantum
infrastructure works in the background. This will be key to illustrate
the dangers and effectiveness of quantum malware on a standard quantum
communications protocol, since Alice and Bob may think that their
automated systems are working and it may take a while for them to
realize that something is wrong, which can typically occur if Alice
sends one of four messages repeatedly or, alternatively, if the protocol
is used multiple times to send longer messages.

Let us address, first, the process for each pattern without the malware,
thus reviewing the standard superdense coding protocol and, afterwards,
discuss how Eve can disrupt the communications circuit with a unitary
gate attack without eavesdropping. Figure 1 shows the four circuits
used for the standard superdense coding, the code in each case is
divided into three sections, the first is the entanglement source
section, the second is Alice\textquoteright s system\textquoteright s
automated operations section and the third is Bob\textquoteright s
system\textquoteright s automated operations. 

\begin{figure}[H]
\begin{centering}
\includegraphics[scale=0.6]{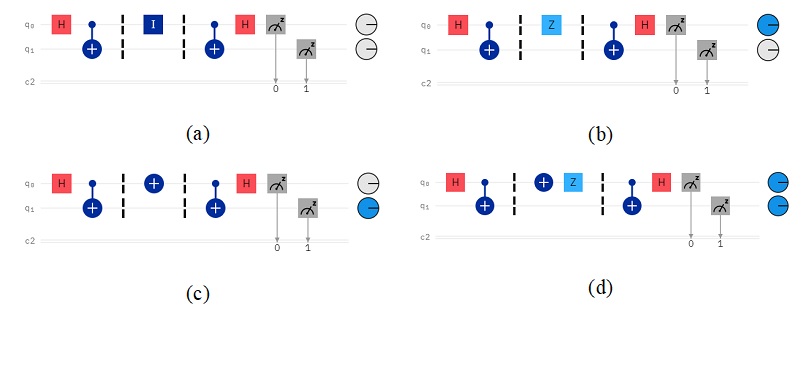}
\par\end{centering}
\caption{Superdense coding circuits, for sending: (a) 00, (b) 01, (c) 10, (d)
11.}
\end{figure}

It should be stressed that we are using the same convention in terms
of notation as IBM\textquoteright s, in order to make it easier to
compare with the experimental implementation, in this notation, when
reading from left to right, Alice\textquoteright s register comes
second while Bob\textquoteright s comes first so that the string is
read $q_{1}q_{0}$, which leads, for the strings 01 and 10, to a replacement
of the stantard protocol\textquoteright s circuits, for which the
notation is the reverse $q_{0}q_{1}$ (Zygelman, 2018).

In each case, if Alice types in the message that is represented by
a string \emph{xy}, with $x,y\in\left\{ 0,1\right\} $, then, the
automated quantum communications system applies the necessary computations
on Alice\textquoteright s qubit and then sends it to Bob, whose system
automatically applies a CNOT gate followed by a Hadamard transform,
measures the two qubits, extracts the corresponding measured string
and translates the message from machine language (string of two bits)
into natural language, showing Alice\textquoteright s high-level message
to Bob.

The following table shows the simulation of the above four circuits
on IBM\textquoteright s quantum computer ibmqx2.

\begin{table}[H]
\begin{centering}
\begin{tabular}{|c|c|c|c|c|}
\hline 
\multirow{2}{*}{Observed String} & \multicolumn{4}{c|}{Intended String}\tabularnewline
\cline{2-5} \cline{3-5} \cline{4-5} \cline{5-5} 
 & 00 & 01 & 10 & 11\tabularnewline
\hline 
\hline 
00 & 91.7\% & 9.3\% & 5.7\% & 0.8\%\tabularnewline
\hline 
01 & 1.3\% & 84.6\% & 1.5\% & 5.1\%\tabularnewline
\hline 
10 & 5.9\% & 1.4\% & 89.7\% & 7.9\%\tabularnewline
\hline 
11 & 1.1\% & 4.7\% & 3.1\% & 86.2\%\tabularnewline
\hline 
\end{tabular}
\par\end{centering}
\caption{Simulation of figure 1\textquoteright s superdense coding protocol
on the ibmqx2 quantum device, with 1000 runs used for each experiment.}

\end{table}

We can see that the simulation on an actual quantum computer contains
some noise, but the intended result holds for the various circuits
with a more than 80\% frequency. The above results constitute a review
and an experimental trial of the superdense coding protocol on IBM\textquoteright s
quantum devices, which can, in this case, be used as simulators of
low-noise quantum communications networks.

Let, now, $\hat{C}_{xy}$, with $x,y\in\left\{ 0,1\right\} $, denote
the sequence of operations for the non-hacked superdense coding, including
the entanglement source operations, followed by Alice\textquoteright s
system\textquoteright s automated operations and then by Bob\textquoteright s
system\textquoteright s automated operations, leading to the final
string \emph{xy}, so that $\hat{C}_{xy}\left|00\right\rangle =\left|xy\right\rangle $.

Malware that changes the automated gate definition, will change the
code for the protocol at one or more of the nodes in the communications
network such that the chains $\hat{C}_{xy}$ are replaced by corresponding
hacked chains $\hat{C}_{xy}^{Hacked}$, which can have one or more
additional unitary operators applied by the automated protocol at
one or more compromised systems.

Bijection attacks, the first type of cyberattacks that we consider,
change the automated gate sequence definition software, so that if
\emph{xy} is the original correct string for the superdense coding
protocol, then we get the recoding:

\begin{equation}
\hat{C}_{xy}^{Hacked}\left|00\right\rangle =e^{i\phi_{xy}}\left|f(xy)\right\rangle 
\end{equation}
where the function $f:\left\{ 00,01,10,11\right\} \rightarrow\left\{ 00,01,10,11\right\} $
is a bijection and $\phi_{xy}$ is an arbitrary phase. Considering
the above equation, it follows that the bijection attacks lead to
a bijective mapping of the projector set $\left\{ \left|xy\right\rangle \left\langle xy\right|:x,y=0,1\right\} $
onto itself, this is due to the fact that the global phase present
in the state vectors disappears for each transformed density under
$\hat{C}_{xy}^{Hacked}$, so that, given a hacked protocol $\hat{C}_{xy}^{Hacked}$
there is a whole family of equivalent hacks all consistent with the
same bijection \emph{f} but that only differ by a U(1) group transformation,
this is an important symmetry for this type of cyberattack.

The second type of attack that we consider uses unitary gates for
scrambling the measurement results, which means that instead of a
computation of a bijective mapping on the main computational basis'
projector set, we get, for each activation of the protocol for a string
\emph{xy}, a superposition state of the form:

\begin{equation}
\hat{C}_{xy}^{Hacked}\left|00\right\rangle =\sum_{w,z\in\{0,1\}}\psi_{xy}(wz)\left|wz\right\rangle 
\end{equation}

Upon measurement, Bob\textquoteright s system will get a random output
with probabilities given by the squared amplitudes $\left|\psi_{xy}(wz)\right|^{2}$,
this also means that the correct output for the superdense coding
will occur with a probability of $\left|\psi_{xy}(xy)\right|^{2}$.

Bijection and unitary scrambling attacks can be uncovered with multiple
uses of the protocol using repeated pre-established test messages,
or when sending of longer strings or, even, under a direct communication
between Alice and Bob which may allow the users to detect a mismatching
between intended messages and received messages. However, in noisy
channels (even with low noise) this randomness may be difficult to
distinguish from a quantum hardware malfunction or strong environmental
noise fluctuations that may have corrupted the communications network.
We now address these two types of attacks, starting with bijection
attacks.

\subsection{Bijection Attacks}

To illustrate bijection attacks, let us assume that Eve has managed
to get malware installed on Alice\textquoteright s system so that
the automated superdense coding protocol is modified, another possibility
would be to attack Bob\textquoteright s system or even the entanglement
source, for the sake of illustration of bijection attack profiles,
we focus first on Alice\textquoteright s system and proceed from there
to address other target nodes.

Now, in this first example, Eve\textquoteright s malware interferes
with Alice\textquoteright s automated coding software, without Alice
knowing that she has been hacked. Under Eve\textquoteright s malware,
a quantum unitary gate or a sequence of unitary gates are, in this
case, applied to Alice\textquoteright s protocol each time it is used,
independently of the typed message, so that no eavesdropping is needed.

We will be considering here the effect of elementary gates in disrupting
superdense coding, so that we will be working, for now, with a single
gate operation. Since the hack takes the form of a single unitary
gate, Eve is not eavesdropping on Alice and Bob, as stated before,
indeed, the automated quantum operation introduced by the malware
is the same whatever the sequence of gates implemented on Alice\textquoteright s
side, the protocol is, in this way, disrupted without Eve having to
eavesdrop on Alice and Bob.

There are several possible unitary gate insertion versions, where
the malware can operate to change Alice system\textquoteright s software-defined
computational circuit. In the case of quantum communications protocols,
where the software automatically changes the circuit in accordance
with the message being sent, as is the case that we are addressing,
it is easier for the malware to either operate at the beginning or
at the end of a system\textquoteright s computational chain, that
is, at the beginning or at the end of the computing system\textquoteright s
automated operations, in this case, the superdense coding protocol
operates normally but there is either an additional final operation,
always the same, that transforms the qubit being sent, or an initial
operation that changes the qubit before employing Alice\textquoteright s
automated operations.

Whether operating at the end or at the beginning, since the operations
are the same, Eve is able to disrupt the superdense coding without
having to eavesdrop on it, therefore, independently of the message
sent by Alice. Figure 2 exemplifies the circuits with an \emph{X}
gate malware attack on the quantum circuit, where the gate always
operates at the end of Alice\textquoteright s system\textquoteright s
quantum computations, while figure 3 shows the operation of the malware
at the beginning of Alice system\textquoteright s computations, both
figures illustrate how the hack works for each of the alternative
automated gate sequences that characterize the superdense coding communications
protocol.

\begin{figure}[H]
\centering{}\includegraphics[scale=0.6]{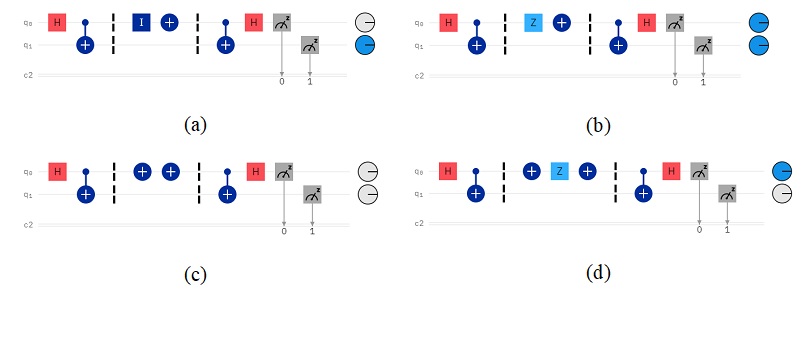}\caption{Hacked superdense coding circuits with \emph{X} gate at the end of
Alice\textquoteright s system\textquoteright s computational chain.}
\end{figure}

\begin{figure}[H]
\begin{centering}
\includegraphics[scale=0.6]{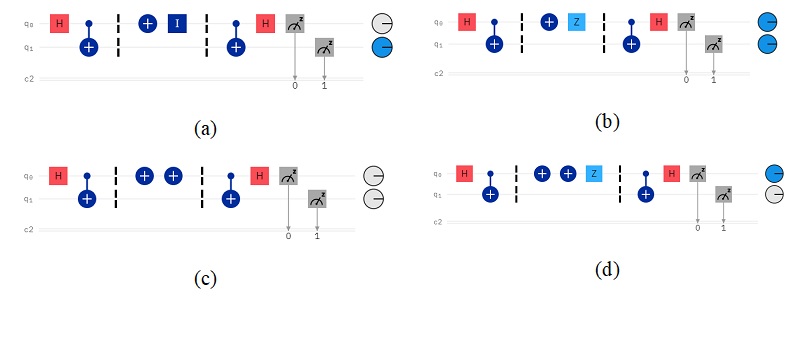}
\par\end{centering}
\caption{Hacked superdense coding circuits with \emph{X} gate at the beginning
of Alice system\textquoteright s computational chain.}
\end{figure}

In table 2, we show the results on the final state vectors from figures
2 and 3\textquoteright s hacked circuits, for the noise-free communications
network. As is visible from the table, the hack works so that in each
case the final state vector differs from the superdense coding intended
result. From the state vectors shown in table 2, assuming a noiseless
communications network, the final measured string when the \emph{X}
gate is automatically applied at the beginning of Alice\textquoteright s
system\textquoteright s automated operations will coincide with the
measured string when the gate is applied at the end, since each final
state vector with the gate applied at the beginning coincides with
the state vector with the gate applied at the end, except in two cases
where there is a global negative phase difference, which does not
affect the final quantum measurement.

\begin{table}[H]
\begin{centering}
\begin{tabular}{|>{\centering}p{3cm}|>{\centering}p{3cm}|>{\raggedright}p{3cm}|}
\hline 
Intended String & \multirow{1}{3cm}{Final state vector with \emph{X} gate at the end of Alice\textquoteright s
chain} & Final state vector with \emph{X} gate at the beginning of Alice\textquoteright s
chain\tabularnewline
\hline 
\hline 
00 & $\left|10\right\rangle $ & $\left|10\right\rangle $\tabularnewline
\hline 
01 & $-\left|11\right\rangle $ & $\left|11\right\rangle $\tabularnewline
\hline 
10 & $\left|00\right\rangle $ & $\left|00\right\rangle $\tabularnewline
\hline 
11 & $-\left|01\right\rangle $ & $\left|01\right\rangle $\tabularnewline
\hline 
\end{tabular}
\par\end{centering}
\caption{Final state vectors resulting from the \emph{X} gate malware at the
beginning and at the end of Alice\textquoteright s system\textquoteright s
computational chain.}
\end{table}

In the ideal noiseless quantum communications context, through her
hack of Alice\textquoteright s computational chain, Eve is capable
of rotating the final output state vector so that Bob\textquoteright s
system always measures the wrong final string.

From these results, it follows that if we work with the final output
density operator at the end of the protocol, we find that the \emph{X}
gate attack, whether at the beginning or the end of Alice\textquoteright s
circuit, leads to the same remapping of the basis densities that is
one-to-one and onto, therefore, bijective.

The reason why the final density with the malware is the same whether
the \emph{X} gate is applied at the beginning or at the end of Alice\textquoteright s
system\textquoteright s operations is due to the fact that the final
state vectors, with the malware, only differ by a global phase when
the \emph{X} gate is applied, we, thus, have an invariance for the
final hacked densities with respect to the placement of the \emph{X}
gate at the beginning or at the end of Alice\textquoteright s computational
chain, since the final state vectors only differ by a global phase,
which means that, working with the final densities, we get the same
bijection for the measurement results, since the projector set is
the same.

This is an important hacking symmetry, furthermore, Eve can change
Bob\textquoteright s measurement, obtaining the same results in terms
of bijective modification by:
\begin{enumerate}
\item Hacking Alice\textquoteright s system with the \emph{X} gate applied
to Alice\textquoteright s qubit at the beginning of Alice\textquoteright s
chain; 
\item Hacking Alice\textquoteright s system with the \emph{X} gate applied
to Alice\textquoteright s qubit at the end of Alice\textquoteright s
chain;
\item Hacking the entanglement source with the \emph{X} gate applied to
Alice\textquoteright s qubit at the end of the entanglement source\textquoteright s
operations;
\item Hacking Bob\textquoteright s system with the \emph{X} gate applied
to Alice\textquoteright s qubit, before the final CNOT and Hadamard
transform.
\end{enumerate}
In an ideal noiseless quantum communications network, the alternatives
1 and 3 lead to the same final state vectors, the same is true for
the alternatives 2 and 4, now still taking advantage of quantum symmetries,
Eve can also obtain the similar results by hacking Bob\textquoteright s
qubit instead of Alice\textquoteright s. For instance, if Eve employs
an \emph{X} gate at the end of the entanglement source\textquoteright s
operations on Bob\textquoteright s qubit instead of Alice\textquoteright s,
then, Eve obtains the same resulting state vectors as those that hold
for the \emph{X} gate attack on Alice\textquoteright s qubit at the
beginning of Alice\textquoteright s standard superdense coding computational
chain, this is due to symmetry and entanglement, namely, the input
for Alice\textquoteright s standard superdense coding computations
is the same whether one applies an \emph{X} gate on Alice\textquoteright s
qubit or on Bob\textquoteright s qubit before the protocol runs these
computations, indeed, after entanglement, hacking Alice\textquoteright s
qubit, after the two qubit\textquoteright s are entangled in accordance
with equation (1)\textquoteright s symmetric pattern, we get the Bell
state which is fed as (hacked) input for Alice and Bob\textquoteright s
standard superdense coding automated operations:

\begin{equation}
\frac{\left|0\right\rangle \otimes X\left|0\right\rangle +\left|1\right\rangle \otimes X\left|1\right\rangle }{\sqrt{2}}=\frac{\left|01\right\rangle +\left|10\right\rangle }{\sqrt{2}}
\end{equation}

This result holds for the two above discussed hacking alternatives
1 and 3. Now, if Eve is able to insert her malware at the entanglement
source so that, after the source entangled the two qubits in accordance
with the symmetric Bell state in equation (1), and if this malware
attacks Bob\textquoteright s qubit with an \emph{X} gate, then, we
get the same hacked input to the remaining superdense coding protocol\textquoteright s
operations as in equation (4), indeed:

\begin{equation}
\frac{X\left|0\right\rangle \otimes\left|0\right\rangle +X\left|1\right\rangle \otimes\left|1\right\rangle }{\sqrt{2}}=\frac{\left|10\right\rangle +\left|01\right\rangle }{\sqrt{2}}=\frac{\left|01\right\rangle +\left|10\right\rangle }{\sqrt{2}}
\end{equation}

Therefore, Eve can get the same final hacking results by attacking
different systems and qubits, taking a strategic advantage of quantum
entanglement and symmetry.

Another example of bijection attack using a different operator is
the \emph{Z} gate attack. In this case, instead of applying an \emph{X}
gate, Eve\textquoteright s malware introduces a \emph{Z} gate at the
beginning or at the end of Alice\textquoteright s chain. Table 3 shows
the results of the \emph{Z} gate attack on the final state vectors,
which again coincide up to a global phase.

\begin{table}[H]
\begin{centering}
\begin{tabular}{|>{\centering}p{3cm}|>{\centering}p{3cm}|>{\raggedright}p{3cm}|}
\hline 
Intended String & \multirow{1}{3cm}{Final state vector with \emph{Z} gate at the end of Alice\textquoteright s
chain} & Final state vector with \emph{Z} gate at the beginning of Alice\textquoteright s
chain\tabularnewline
\hline 
\hline 
00 & $\left|01\right\rangle $ & \centering{}$\left|01\right\rangle $\tabularnewline
\hline 
01 & $\left|00\right\rangle $ & \centering{}$\left|00\right\rangle $\tabularnewline
\hline 
10 & $\left|11\right\rangle $ & \centering{}$-\left|11\right\rangle $\tabularnewline
\hline 
11 & $\left|10\right\rangle $ & \centering{}$-\left|10\right\rangle $\tabularnewline
\hline 
\end{tabular}
\par\end{centering}
\caption{Final state vectors resulting from the \emph{Z} gate malware at the
beginning and at the end of Alice\textquoteright s system\textquoteright s
computational chain.}
\end{table}

These two malware attacks are examples of the general class of bijection
attacks that remap the algorithm\textquoteright s final densities
with a one-to-one and onto transformation defined on the projector
set $\left\{ \left|xy\right\rangle \left\langle xy\right|:x,y=0,1\right\} $.
Eve\textquoteright s intention, in this case, is to change the final
message in a specific way, so that, in the noiseless communications
circuit, the final measured string will be different from Alice\textquoteright s
intended message.

Now, since these attacks lead to a deterministic invertible remapping
of the intended two bits string onto another two bits string, the
one-to-one and onto nature of this remapping is a weakeness for a
long-term hack, which means that this can only work as a short-term
attack aimed at fooling Alice and Bob\textquoteright s communications
for a short period, in order to disrupt communications. It introduces
a form of man-in-the-middle variant where Eve replaces Alice\textquoteright s
messages in a specific way, and Bob is fooled into believing that
he is receiving Alice\textquoteright s true message while he is in
fact receiving Eve\textquoteright s altered messages.

If Alice and Bob communicate or use a changing preestablished test
message, each time they use the system, then the hack can be uncovered.
However, the test message solution can be overturned depending on
the way it is built, for instance, automated test messages if hacked
as well can lead to a longer period until the hack is found, unless
Alice and Bob communicated directly or use multiple communication
channels.

Bijection attacks can also be useful if the context is not one of
message reading by Bob, but rather one in which we are dealing with
distributed automated networked quantum computation, such that Alice\textquoteright s
system provides an initial state for Bob\textquoteright s system to
perform aditional automated computations on the final outputs coming
from the communications network, in this context, Eve\textquoteright s
malware will alter Bob\textquoteright s further computations in a
controlled way. In this case, the malware can also take a while longer
to be uncovered, depending on the additional computations performed
on the outputs coming from the communications network.

In either case, the above results show a vulnerability of the quantum
superdense coding protocol to bijection attacks. However, once uncovered,
bijection attacks are easy to counter, one can produce a form of quantum
quarantine of the malware, by remapping on Bob\textquoteright s end
the final densities through a unitary rotation that restores the original
protocol, leaving the cybersecurity team ample time to check the network
nodes for the malware. During quarantine, one can even fool Eve into
believing that her attack has not been countered.

Now, while with bijection attacks, Eve wishes to change the measured
output in a controlled way, scrambling attacks are different since
their strategic goal is to use unitary gates to produce a random result
upon measurement, taking advantage of quantum randomness upon measurement,
damaging quantum communications, this quantum communications scrambling
attack through unitary gates becomes more difficult to distinguish
from a communications circuit corrupted by environmental noise, even
though the scrambling attack is unitary. If a quantum communications
circuit has some noise, even if small, then, the cyberattack may increase
the circuit\textquoteright s random output, furthermore, this type
of malware is not as easily quarantined as the previous one.

We will now address examples of unitary scrambling attacks on the
quantum superdense coding and how entanglement and symmetries can
be used to hide the attacked system, thus becoming strategic key drivers
for quantum cyberoperations.

\subsection{Unitary Scrambling Attacks}

Unitary scrambling attacks employ malware to corrupt automatic gate
encoding and take advantage of quantum randomness at the level of
the final measurement in order to disrupt a quantum communications
protocol, masking the cyberattack under the cover of quantum environmental
noise corruption or a quantum hardware malfunction, when, in fact,
the error level of the quantum communications circuit with respect
to the intended output was increased not due to external noise but
rather through a unitary gate that produced a quantum superposition
with amplitudes that lead to significant deviations, upon measurement,
from a final intended output, taking advantage of the usual quantum
randomness upon measurement to give the appearance of rise in communications
channel noise, such that, unaware that the communications network
has been hacked, the participants will get random fluctuations in
the measurement outputs in a way that will no longer match the communications
protocol.

In this case, Eve\textquoteright s hidding of the attack can only
be short lived in the sense that it may not take a long time to find
out that something is wrong with the commmunications network, Eve\textquoteright s
intention here is similar to a DDoS attack, since her endgame is to
disrupt the communications protocol by producing random measurement
results for one or more qubits, the effectiveness of this attack is
enhanced by a few points: on the one hand, such an attack can be difficult
to distinguish from a hardware malfunction, or environmental noise
increase in a noisy communications network, which may lead to an additional
cost from the part of systems\textquoteright{} maintenance to identify
the source of output error, on the other hand, as we will show, by
taking advantage of entanglement and symmetries involved in a quantum
communications protocol, a hacker can attack one system and one qubit
but produce a disruption on another qubit that never passed through
a hacked system, this hacking propagation makes it more difficult
to uncover which system and qubits were actually hacked.

Once a disruption takes place, quantum cybersecurity teams in collaboration
with system\textquoteright s maintenance will have to identify the
disrupted node and check for hardware malfunction or actual malware
operating at the level of the automated gate translation.

An example of such a software attack profile, for the superdense coding
protocol, can be obtained through an \emph{S} gate attack on Alice\textquoteright s
system, this is when Eve installs malware in Alice\textquoteright s
automated gate specification introducing an additional \emph{S} gate
operating either at the end or at the beginning of Alice\textquoteright s
operator chain, for both these cases, as shown in table 4, the attack
produces, at the end of the superdense coding protocol, a superposition
at the level of Alice\textquoteright s qubit, while leaving Bob\textquoteright s
qubit in the correct superdense coding protocol\textquoteright s configuration.

\begin{table}[H]
\begin{raggedright}
\begin{tabular}{|>{\raggedright}p{4cm}|>{\raggedright}p{4cm}|>{\raggedright}p{4cm}|}
\hline 
Intended String & \multirow{1}{4cm}{Final state vector with \emph{S} gate at the end of Alice\textquoteright s
chain} & Final state vector with \emph{S} gate at the beginning of Alice\textquoteright s
chain\tabularnewline
\hline 
\hline 
00 & \centering{}$\left|0\right\rangle \otimes\frac{(1+i)\left|0\right\rangle +(1-i)\left|1\right\rangle }{2}$ & \centering{}$\left|0\right\rangle \otimes\frac{(1+i)\left|0\right\rangle +(1-i)\left|1\right\rangle }{2}$\tabularnewline
\hline 
01 & \centering{}$\left|0\right\rangle \otimes\frac{(1-i)\left|0\right\rangle +(1+i)\left|1\right\rangle }{2}$ & \centering{}$\left|0\right\rangle \otimes\frac{(1-i)\left|0\right\rangle +(1+i)\left|1\right\rangle }{2}$\tabularnewline
\hline 
10 & \centering{}$\left|1\right\rangle \otimes\frac{(1+i)\left|0\right\rangle +(1-i)\left|1\right\rangle }{2}$ & \centering{}$\left|1\right\rangle \otimes\frac{(1+i)\left|0\right\rangle -(1-i)\left|1\right\rangle }{2}$\tabularnewline
\hline 
11 & \centering{}$\left|1\right\rangle \otimes\frac{(1-i)\left|0\right\rangle +(1+i)\left|1\right\rangle }{2}$ & \centering{}$\left|1\right\rangle \otimes\frac{-(1-i)\left|0\right\rangle +(1+i)\left|1\right\rangle }{2}$\tabularnewline
\hline 
\end{tabular}
\par\end{raggedright}
\caption{Final state vectors resulting from the \emph{S} gate attack at the
beginning and at the end of Alice\textquoteright s computational chain.}
\end{table}

Table 5 shows the results from the simulation of the hacked protocol
on the ibmqx2 device resulting from the \emph{S} gate applied at the
beginning of Alice\textquoteright s chain.

\begin{table}[H]
\begin{centering}
\begin{tabular}{|c|c|c|c|c|}
\hline 
\multirow{2}{*}{Observed String} & \multicolumn{4}{c|}{Intended String}\tabularnewline
\cline{2-5} \cline{3-5} \cline{4-5} \cline{5-5} 
 & 00 & 01 & 10 & 11\tabularnewline
\hline 
\hline 
00 & 45.4\% & 47.8\% & 5.7\% & 6.5\%\tabularnewline
\hline 
01 & 49.5\% & 46.6\% & 5.4\% & 5.6\%\tabularnewline
\hline 
10 & 3.3\% & 2.7\% & 40.7\% & 44.4\%\tabularnewline
\hline 
11 & 1.8\% & 2.9\% & 48.2\% & 43.5\%\tabularnewline
\hline 
\end{tabular}
\par\end{centering}
\caption{Simulation of the hacked protocol on ibmqx2, with the \emph{S} gate
operating at the beginning of Alice\textquoteright s computational
chain. In each case, 1000 runs were used.}
\end{table}

The simulations on the ibmqx2 device, which is a low noise quantum
computer, helps to simulate the results from implementing the protocol
on what would be a low noise quantum communications network, in this
case, as shown in table 5, while the final measurement shows that
Bob\textquoteright s qubit is in the correct configuration for most
of the runs of the protocol (94.9\% for the message 00, 94.4\% for
the message 01, 88.9\% for the message 10 and 87.9\% for the message
11), Alice\textquoteright s qubit shows a large deviation from the
protocol\textquoteright s intended output.

Since only the final measurement results are observable for the users,
what the users will see, if they test the circuit with multiple test
messages, is a rise in the error rate of Alice\textquoteright s qubit,
indeed, in the example of table 5\textquoteright s simulations, in
the case of the intended string 00, Alice\textquoteright s qubit was
measured with the correct output for 48.7\% of the runs, while in
the case of the intended string 01, it was measured with the correct
output for 49.5\% of the runs, for the intended string 10, in turn,
we get the correct output in 46.4\% of the runs and, finally, for
the intended string 11, we get the correct output in 49.1\% of the
runs.

Considering the intended non-hacked protocol\textquoteright s outputs,
table 6 shows the error rates for each qubit with respect to the superdense
coding intended outputs, these results from simulating the hacked
protocol on ibmqx2 help illustrate what would be the consequences
of hacking a low noise quantum communications network, which will
be useful in the forensic considerations that we now address.

\begin{table}[H]
\begin{centering}
\begin{tabular}{|>{\centering}p{3cm}|>{\centering}p{3cm}|>{\raggedright}p{3cm}|}
\hline 
Intended String & \multirow{1}{3cm}{Error rate for Alice\textquoteright s qubit} & Error rate for Bob\textquoteright s qubit\tabularnewline
\hline 
\hline 
00 & 51.3\% & \centering{}5.1\%\tabularnewline
\hline 
01 & 50.5\% & \centering{}5.6\%\tabularnewline
\hline 
10 & 53.6\% & \centering{}11.1\%\tabularnewline
\hline 
11 & 50.9\% & \centering{}12.1\%\tabularnewline
\hline 
\end{tabular}
\par\end{centering}
\caption{Error rates between the measured string and the superdense coding
intended outputs, obtained from table 5\textquoteright s experimental
frequencies.}
\end{table}

In tables 5 and 6\textquoteright s simulations, the error rate for
Bob is within the boundaries of standard random fluctuations in what
would be a low noise quantum communications network, with the minimum
error rate for Bob being of 5.1\% and the maximum error rate being
of 12.1\%, however, for Alice, we get the increase in the error rate
to a minimum level of 50.5\% and a maximum level of 53.6\%. From a
quantum cybersecurity forensics standpoint this would be a warning
signal on possible hardware malfunction, a strong environmental disruption
or, alternatively, a cyberattack aimed at disrupting Alice\textquoteright s
qubit.

Since the disrupted qubit is Alice\textquoteright s, in the case of
a quantum cybersecurity investigation on a possible software hacking
of automated gate definition, in the context of a unitary scrambling
attack, the investigation might be led to infer a possible hacking
targeting directly Alice\textquoteright s qubit, with the probable
attack node having been Alice\textquoteright s software for the superdense
coding protocol\textquoteright s automation, however, this inference
does not necessarily hold for a quantum cybersecurity context as we
now show.

In the context of quantum communications hacking, by taking strategic
advantage of a non-hacked protocol\textquoteright s employment of
entanglement and symmetry, a hacker may be able to disrupt a specific
qubit that has been sent through and processed by a specific (sub)network
by hacking another entangled qubit that has followed another route
of nodes, in this case, even if the two qubits follow separate paths,
because they are entangled, a hacker can change one qubit by attacking
another qubit along its quantum processing route in the communications
network. This characteristic of quantum communications, which incorporates
quantum networked computation, makes it more difficult for quantum
cybersecurity forensics to find the attacked nodes and qubits, especially
in large quantum communications networks.

In our example, the network is small containing only three nodes:
the entanglement source, Alice's system and Bob's system, however,
as we now show, Alice\textquoteright s qubit can be disrupted with
a similar pattern as that of tables 4 to 6 by operating on Bob\textquoteright s
qubit at the entanglement source or at Bob\textquoteright s system.

Let us, first, consider the scenario in which Eve is able to insert
the malware at the entanglement source with an \emph{S} gate attack
on Bob\textquoteright s qubit where the \emph{S} gate operates at
the end of the entanglement source\textquoteright s superdense coding
computational chain and before the qubit is sent to Bob. In table
7, we show the results from simulating this attack on the ibmqx2 quantum
device, these results show a very similar profile to that of table
5, indeed, the error with respect to the non-hacked superdense coding
protocol\textquoteright s output is raised for Alice\textquoteright s
qubit rather than Bob\textquoteright s, even though, this time, Eve\textquoteright s
malware led to the modified circuit to operate solely on Bob\textquoteright s
qubit.

\begin{table}[H]
\begin{centering}
\begin{tabular}{|c|c|c|c|c|}
\hline 
\multirow{2}{*}{Observed String} & \multicolumn{4}{c|}{Intended String}\tabularnewline
\cline{2-5} \cline{3-5} \cline{4-5} \cline{5-5} 
 & 00 & 01 & 10 & 11\tabularnewline
\hline 
\hline 
00 & 41.1\% & 47.6\% & 6.0\% & 5.6\%\tabularnewline
\hline 
01 & 50.6\% & 45.0\% & 7.8\% & 8.5\%\tabularnewline
\hline 
10 & 4.5\% & 2.0\% & 41.4\% & 40.8\%\tabularnewline
\hline 
11 & 3.8\% & 5.4\% & 44.8\% & 45.1\%\tabularnewline
\hline 
\end{tabular}
\par\end{centering}
\caption{Simulation of the hacked protocol on ibmqx2, with the \emph{S} gate
operating on Bob\textquoteright s qubit at the end of the entanglement
source\textquoteright s computational chain. In each case, 1000 runs
were used.}
\end{table}

In table 8, we show the error rates with respect to the superdense
coding protocol for each qubit, which reinforces the above results,
showing that even though Eve\textquoteright s malware operated on
Bob\textquoteright s qubit, it was Alice\textquoteright s qubit that
was \textquotedblleft scrambled\textquotedblright .

\begin{table}[H]
\begin{centering}
\begin{tabular}{|>{\centering}p{3cm}|>{\centering}p{3cm}|>{\raggedright}p{3cm}|}
\hline 
Intended String & \multirow{1}{3cm}{Error rate for Alice\textquoteright s qubit} & Error rate for Bob\textquoteright s qubit\tabularnewline
\hline 
\hline 
00 & 54.4\% & \centering{}8.3\%\tabularnewline
\hline 
01 & 49.6\% & \centering{}7.4\%\tabularnewline
\hline 
10 & 52.6\% & \centering{}13.8\%\tabularnewline
\hline 
11 & 46.4\% & \centering{}14.1\%\tabularnewline
\hline 
\end{tabular}
\par\end{centering}
\caption{Error rates between the measured string and the superdense coding
intended outputs, obtained for table 7\textquoteright s experimental
frequencies.}
\end{table}

There is a reason for the results to be similar, indeed, the final
state vectors from an \emph{S} gate attack on Bob\textquoteright s
qubit at the end of the entanglement source\textquoteright s operations
coincide with the final state vectors for an \emph{S} gate attack
on Alice\textquoteright s qubit at the beginning of Alice\textquoteright s
operator chain. These results are specific to quantum communications,
and they are linked to entanglement and symmetry, namely, the non-hacked
entanglement source\textquoteright s operations have the following
structure:

\begin{equation}
\hat{C}_{Ent}=\left(I\otimes\left|0\right\rangle \left\langle 0\right|+X\otimes\left|1\right\rangle \left\langle 1\right|\right)\left(I\otimes H\right)
\end{equation}

Considering the initial state $\left|00\right\rangle $, we get equation
(1)\textquoteright s symmetric Bell state for Bob and Alice:

\begin{equation}
\hat{C}_{Ent}\left|00\right\rangle =\frac{\left|00\right\rangle +\left|11\right\rangle }{\sqrt{2}}
\end{equation}

Now, the second qubit is sent to Alice while the first is sent to
Bob. In the case where Alice\textquoteright s system is hacked with
an \emph{S} gate applied at the beginning of the superdense coding
operations for Alice, the state vector after the hack and before Alice\textquoteright s
standard superdense coding operations becomes:

\begin{equation}
\frac{\left|0\right\rangle \otimes S\left|0\right\rangle +\left|1\right\rangle \otimes S\left|1\right\rangle }{\sqrt{2}}=\frac{\left|00\right\rangle +i\left|11\right\rangle }{\sqrt{2}}
\end{equation}

In this way, because Alice and Bob\textquoteright s qubits are entangled,
the S gate attack has changed the entangled state before running the
standard superdense coding protocol\textquoteright s operations for
Alice, adding a phase to the branch $\left|11\right\rangle $.

On the other hand, if, instead of attacking Alice\textquoteright s
qubit, the hack is on Bob\textquoteright s but at the end of the entanglement
source\textquoteright s operations, then, the hacked entanglement
source\textquoteright s computational chain is given by:

\begin{equation}
\hat{C}_{Ent}^{Hacked}=\left(S\otimes I\right)\hat{C}_{Ent}
\end{equation}

Which leads to the following entanglement pattern at the end of the
entanglement source\textquoteright s computations:

\begin{equation}
\hat{C}_{Ent}^{Hacked}\left|00\right\rangle =\frac{S\left|0\right\rangle \otimes\left|0\right\rangle +S\left|1\right\rangle \otimes\left|1\right\rangle }{\sqrt{2}}=\frac{\left|00\right\rangle +i\left|11\right\rangle }{\sqrt{2}}
\end{equation}

This is the same output that is obtained when Eve attacks Alice\textquoteright s
qubit, instead of Bob\textquoteright s, since the remaining operations
coincide for both hacks, when Eve hacks Bob\textquoteright s qubit
at the end of the entanglement source with an \emph{S} gate attack
she gets the same state vectors before Alice\textquoteright s standard
superdense coding operations than when she hacks Alice\textquoteright s
system adding the \emph{S} gate at the beginning of Alice\textquoteright s
system\textquoteright s standard superdense coding operations, this
is due to the symmetry of the Bell state that results from the entanglement
source\textquoteright s operations.

Now, one might be led to infer that an attack on Bob\textquoteright s
qubit leading to a scrambling of Alice\textquoteright s qubit, instead
of Bob\textquoteright s could only come from a hack at the entanglement
source, however, this is not the case, indeed, Eve can attack Bob\textquoteright s
qubit with the scrambling occurring for Alice\textquoteright s qubit
by hacking Bob\textquoteright s system and changing Bob\textquoteright s
operators. In this case, if, instead of attacking the entanglement
source or Alice\textquoteright s system, Eve is able to get the malware
into Bob\textquoteright s system, changing Bob\textquoteright s system\textquoteright s
automated gate definition, so that an \emph{S} gate is applied to
Bob\textquoteright s qubit at the beginning of Bob\textquoteright s
system\textquoteright s superdense coding operations, then, as shown
in table 9, when compared with table 4\textquoteright s results, we
also find that the hacked protocol leads to final state vectors that
coincide with those that hold when Alice\textquoteright s system is
hacked with the \emph{S} gate added at the beginning of Alice\textquoteright s
chain, attacking Alice\textquoteright s qubit.

\begin{table}[H]
\begin{centering}
\begin{tabular}{|>{\raggedright}p{4cm}|>{\raggedright}p{4cm}|}
\hline 
Intended String & Final state vector with \emph{S} gate at the beginning of Bob\textquoteright s
chain\tabularnewline
\hline 
\hline 
00 & \centering{}$\left|0\right\rangle \otimes\frac{(1+i)\left|0\right\rangle +(1-i)\left|1\right\rangle }{2}$\tabularnewline
\hline 
01 & \centering{}$\left|0\right\rangle \otimes\frac{(1-i)\left|0\right\rangle +(1+i)\left|1\right\rangle }{2}$\tabularnewline
\hline 
10 & \centering{}$\left|1\right\rangle \otimes\frac{(1+i)\left|0\right\rangle -(1-i)\left|1\right\rangle }{2}$\tabularnewline
\hline 
11 & \centering{}$\left|1\right\rangle \otimes\frac{-(1-i)\left|0\right\rangle +(1+i)\left|1\right\rangle }{2}$\tabularnewline
\hline 
\end{tabular}
\par\end{centering}
\caption{Final vectors resulting from the \emph{S} gate malware at the beginning
of Bob\textquoteright s chain.}
\end{table}

By hacking Bob\textquoteright s qubit, this time at Bob\textquoteright s
system, we get the same profile that we would get from attacking Alice\textquoteright s
qubit directly at the beginning of Alice\textquoteright s chain as
well as the same profile from hacking Bob\textquoteright s qubit at
the end of the entanglement chain. As in the bijection attacks, we
get a symmetry profile for hacking, by which different hacks lead
to the same final result, this is strategically advantageous to the
hacker, making it harder for quantum cybersecurity to find out which
network node and qubit were attacked. In table 10 we show the results
from simulating this last hacked protocol on ibmqx2.

\begin{table}[H]
\begin{centering}
\begin{tabular}{|c|c|c|c|c|}
\hline 
\multirow{2}{*}{Observed String} & \multicolumn{4}{c|}{Intended String}\tabularnewline
\cline{2-5} \cline{3-5} \cline{4-5} \cline{5-5} 
 & 00 & 01 & 10 & 11\tabularnewline
\hline 
\hline 
00 & 46.1\% & 49.3\% & 7.9\% & 6.0\%\tabularnewline
\hline 
01 & 45.8\% & 44.1\% & 8.1\% & 8.1\%\tabularnewline
\hline 
10 & 5.6\% & 3.0\% & 43.4\% & 42.1\%\tabularnewline
\hline 
11 & 2.5\% & 3.6\% & 40.6\% & 43.8\%\tabularnewline
\hline 
\end{tabular}
\par\end{centering}
\caption{Simulation of the hacked protocol on ibmqx2, with the \emph{S} gate
operating on Bob\textquoteright s qubit at the beginning of Bob\textquoteright s
operator chain. In each case, 1000 runs were used.}
\end{table}

Looking at the final frequencies of measured strings and comparing
with those of tables 5 and 7 we find that we cannot easily distinguish
between the different attacks, indeed, from a quantum cybersecurity
forensics standpoint, from the evidence of disruption on Alice\textquoteright s
qubit, and having ruled out hardware malfunction or rise in environmental
noise, the evidence is favorable to a hack. In the case of a software
hacking targeting automated gate definition in the communications
protocol, the only thing that can be inferred is the hypothesis that
at least one of the communications network nodes and qubit have been
attacked, but which node and qubit one cannot ascertain without inspecting
each one for software changes, this is the main strategic advantage
of quantum hacking symmetries that can be exploited by a quantum hacker.

As another example, for instance, Eve can obtain a similar final frequency
profile upon measurement hacking Bob\textquoteright s system and Alice\textquoteright s
qubit with a Hadamard gate applied to Alice\textquoteright s qubit
at the end of Bob\textquoteright s circuit, in table 11 we show the
simulation on ibmqx2 that results from applying this alternative hack,
in this case, we have a similar measurement pattern but Eve used a
different gate (a Hadamard gate rather than an \emph{S} gate) and
operated on Alice\textquoteright s qubit at the end of Bob\textquoteright s
operator chain.

\begin{table}[H]
\begin{centering}
\begin{tabular}{|c|c|c|c|c|}
\hline 
\multirow{2}{*}{Observed String} & \multicolumn{4}{c|}{Intended String}\tabularnewline
\cline{2-5} \cline{3-5} \cline{4-5} \cline{5-5} 
 & 00 & 01 & 10 & 11\tabularnewline
\hline 
\hline 
00 & 47.9\% & 41.7\% & 5.9\% & 9.3\%\tabularnewline
\hline 
01 & 47.4\% & 50.9\% & 7.9\% & 8.2\%\tabularnewline
\hline 
10 & 3.1\% & 4.9\% & 42.4\% & 42.2\%\tabularnewline
\hline 
11 & 1.6\% & 2.5\% & 43.8\% & 40.3\%\tabularnewline
\hline 
\end{tabular}
\par\end{centering}
\caption{Simulation of the hacked protocol on ibmqx2, with the \emph{H} gate
operating on Alice\textquoteright s qubit at the end of Bob\textquoteright s
operator chain. In each case, 1000 runs were used.}
\end{table}

These results show that different operators, attack nodes and qubits
can all lead to similar quantum measurement frequency profiles, making
it difficult to infer which node and qubit were actually attacked
and what type of modification was actually used. The fact that attacking
one qubit can lead to a disruption in another qubit, makes scrambling
attacks on larger networks an effective security threat for quantum
communications.

The implication of these results for cybersecurity is significant,
since it provides an example of how one can hide a hack by attacking
one qubit but in the end the disrupted qubit will be another one,
that was never directly operated upon through the installed the malware,
this is a specific feature of cyberattacks on quantum networks and
communications\textquoteright{} automated protocols: by strategically
taking advantage of entanglement and symmetries for hacking quantum
networked computation and communications makes it difficult to determine
which systems and qubits were actually attacked.

In this sense, while a DDoS attack may be easier to detect as such,
a quantum unitary scrambling attack, by compromising automation software
with malware, may take a longer time to be uncovered, not only due
to the need to rule out hardware malfunction or external factors like
environmental noise that may have corrupted the quantum communications
network, but also due to the the ability to propagate hacks from one
attacked qubit to another, leaving the first attacked qubit unchanged
at the final measurement, as exemplified above.

In the above example, since Alice\textquoteright s qubit was the one
showing the increase in error with respect to the superdense coding
protocol\textquoteright s unhacked output, while Bob\textquoteright s
qubit followed the protocol with low noise deviations, one might waste
time by first trying to find out if there was a hardware malfunction
on each network node where computations were performed on Alice\textquoteright s
qubit, and then, ruling out a physical source for the disruption,
one might first try to find out if Alice\textquoteright s gate definition
software for the protocol was hacked with installed malware. However,
if no such malware is found on Alice\textquoteright s system, then,
one would still have to check the entanglement source and Bob\textquoteright s
system for possible malware, since Eve might have instead hacked Bob\textquoteright s
system or the entanglement source, attacking either Alice\textquoteright s
qubit or Bob\textquoteright s qubit, leading to the same disruption
pattern on Alice\textquoteright s qubit, while leaving Bob\textquoteright s
qubit with the correct main superdense coding pattern. From the observed
frequencies, one cannot know exactly which node or qubit of the entangled
pair was in fact attacked.

On a small network, such as the above, the forensic process may not
take too long, however, when dealing with larger networks, it may
indeed take a while to identify which systems and qubits were actually
attacked, from the moment in which the attack is identified as such,
since, by taking advantage of entanglement and symmetries, a hacker
can mask the attacked nodes and qubits in such a way that upon measurement,
the final disrupted qubit may not have ever entered an attacked system,
the hack is effectively propagated from one qubit to another, due
to the manipulation of entanglement and symmetries, while leaving
the initial attacked qubit unchanged, as exemplified above for the
superdense coding.

At this point, quantum cyberforensics becomes even more complex, since
not only can different unitary gates lead to a similar measured profile
but also the attacked systems can be masked through hack propagation,
so that it becomes difficult to ascertain, especially in circuits
that use quantum entanglement and symmetries, which endpoint was hacked.
This gives a strategic and tactical advantage for short term quantum
communications disruptions, coming from unitary scrambling attacks
using unitary gates to induce deviations from automated communications
protocols\textquoteright{} intended outputs, producing random results
upon measurement rather than the protocol\textquoteright s intended
outputs, taking advantage of quantum randomness upon measurement.

Since the amplitudes are unobservable, only the final frequency distribution
is available and since one cannot find out which systems or qubits
were attacked just from observing the final measurement frequency
distribution, as shown above, with different possible quantum malware
attacks leading to the same final result, including even, also, the
same final quantum amplitudes, a quantum measurement frequency analysis
does not provide for a forensic basis on which to decide which networked
systems and qubits were hacked. Only by looking for malware at each
communication node can one find out where the malware was inserted.

In large quantum communications networks, with communications protocol
automation and multiple possible nodes attacked by malware installed
to change automatic gate definition, unitary scrambling attacks can
effectively lead to a propagation of hacks and disrupt a quantum communications
network as well as distributed networked quantum computation, due
to the exploitation of entanglement patterns and symmetries involved
in quantum communications protocols for the process of quantum information
transmission and processing, raising the error rates with respect
to intended outputs and making it very difficult to find, from the
final distribution, which nodes were attacked.

Eve can also attack multiple nodes, and produce disruptions on multiple
qubits upon final measurement. As an example of malware attacking
the automated protocol on two systems (the entanglement source and
Alice), table 12 shows the results from a simulation of a two-point
hack, when the intended string for the non-hacked circuit is 00, with
an \emph{S} gate attack at the end of Alice\textquoteright s computational
chain and a $\sqrt{X}$ gate attack applied to Bob\textquoteright s
qubit at the end of the entanglement source\textquoteright s computational
chain, as can be seen, the final frequencies seem to be close to an
equiprobable distribution over the different four alternative strings.
Indeed, the pattern observed in the simulations both for the qasm\_simulator
and the actual physical device (ibmqx2), shown in table 12, match
well the theoretical probabilities, which, in this case are 0.25 for
each string.

\begin{table}[H]
\begin{centering}
\begin{tabular}{|>{\centering}p{3cm}|>{\centering}p{3cm}|>{\raggedright}p{3cm}|}
\hline 
Intended String & \multirow{1}{3cm}{\centering{}Hacked Circuit (qasm\_simulator) } & \centering{}Hacked Circuit (ibmqx2) \tabularnewline
\hline 
\hline 
00 & \centering{}24.5\% & \centering{}22.8\%\tabularnewline
\hline 
01 & \centering{}23.7\% & \centering{}25.3\%\tabularnewline
\hline 
10 & \centering{}25.1\% & \centering{}24.4\%\tabularnewline
\hline 
11 & \centering{}26.7\% & \centering{}27.5\%\tabularnewline
\hline 
\end{tabular}
\par\end{centering}
\caption{Simulation of the hacked circuit, on ibmqx2 and ibm\_qasm\_simulator,
when the superdense coding protocol is applied for 00 message, with
a $\sqrt{X}$ gate applied on Bob\textquoteright s qubit at the end
of the entanglement source\textquoteright s operations and the \emph{S}
gate malware at the end of Alice\textquoteright s chain. In each case
1000 runs were used.}
\end{table}

In figure 4 we show the circuits for this attack, that we call $\sqrt{X}+S$
gate attack, for each alternative superdense coding circuit that makes
up the superdense coding protocol.

\begin{figure}[H]
\begin{centering}
\includegraphics[scale=0.6]{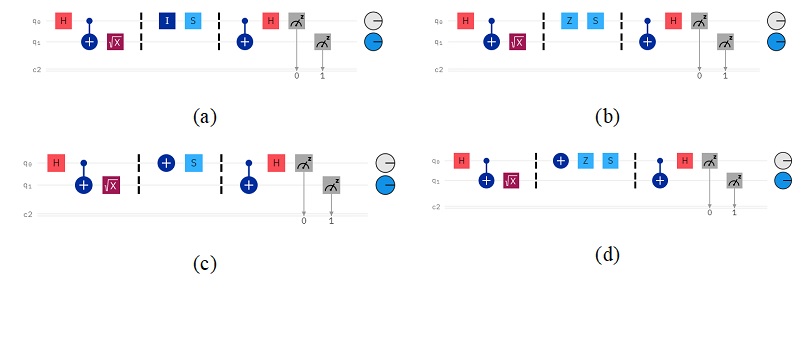}
\par\end{centering}
\caption{Hacked superdense coding circuits with $\sqrt{X}$ gate attack at
the end of the entanglement source operations on Bob\textquoteright s
qubit and \emph{S} gate attack at the end of Alice\textquoteright s
computational chain.}
\end{figure}

In each of the above circuits, upon measurement, we get an approximately
25\% probability distribution over the four alternative strings, which
comes from the fact that the final state vectors lead to an equiprobable
distribution over the alternative values, a result that is due to
the final state vector superposition configuration for each alternative
activation of the superdense coding protocol, as we show in table
13.

\begin{table}[H]
\begin{centering}
\begin{tabular}{|>{\centering}p{3cm}|>{\raggedright}p{6cm}|}
\hline 
Intended String & \multirow{1}{6cm}{$\sqrt{X}+S$ gate attack state vectors}\tabularnewline
\hline 
\hline 
00 & \centering{}$\frac{i\left|00\right\rangle +\left|01\right\rangle +\left|10\right\rangle -i\left|11\right\rangle }{2}$\tabularnewline
\hline 
01 & \centering{}$\frac{\left|00\right\rangle +i\left|01\right\rangle -i\left|10\right\rangle +\left|11\right\rangle }{2}$\tabularnewline
\hline 
10 & \centering{}$\frac{\left|00\right\rangle -i\left|01\right\rangle +i\left|10\right\rangle +\left|11\right\rangle }{2}$\tabularnewline
\hline 
11 & \centering{}$\frac{-i\left|00\right\rangle +\left|01\right\rangle +\left|10\right\rangle +i\left|11\right\rangle }{2}$\tabularnewline
\hline 
\end{tabular}
\par\end{centering}
\caption{Final vectors resulting for each circuit in figure 4.}
\end{table}

This last result shows how a hacker can disrupt several qubits simultaneously,
in this case, Eve scrambles both Alice and Bob\textquoteright s qubits.

\section{Conclusions and Discussion}

We showed that superdense coding, with the automated communications
protocol managed by software, is vulnerable to cyberattacks that come
from malware targeting the automation software and changing the automated
unitary gate definition, effectively changing the communications protocol.
When communications are operating automatically, with the users not
looking at the actual quantum gate sequence, this can lead to disrupted
quantum communications and, also, disrupted quantum networked computation.

We analyzed two types of attacks that can result from malware changing
automated gate sequence definition in communications protocols, using
superdense coding as an example. Both attack types involve unitary
gate insertion at one or more communication nodes, changing the quantum
computations performed at those nodes, however, the two attack types
differ in the final profile and strategic intention.

The first attack type, which we called a bijection attack, works as
a bijective remapping of the final transmitted message for each alternative
message binary coding that Alice can send. This attack allows for
Eve to change the message that Bob receives in a controlled way.

The second attack type, that we called unitary scrambling attack,
operates so that the communications are scrambled, which means that
upon measurement one or more qubits exhibit random fluctuations. This
attack is more versatile in the sense that it can be employed for
different operational profiles, namely, it can be used to produce
final random results, upon measurement, for one, several or even all
qubits.

To produce final random results upon measurement can be difficult
to distinguish from hardware malfunction or sudden environmental noise
increase, demanding an extra effort on the part of quantum cybersecurity
forensics to identify the source of the disruption. Another advantage
of producing random results at one or more attack target qubits is
such that the hacker can exploit entanglement and symmetry, from a
hacker\textquoteright s standpoint, this is one of its main effectivenesses
since one can disrupt a target qubit by operating on another qubit
of an entangled pair, thus propagating the hack from one qubit to
another, making it difficult to ascertain which communications\textquoteright{}
node and qubit were attacked.

From a quantum cybersecurity standpoint, the results of the present
work indicate the need to:
\begin{enumerate}
\item Develop research into quantum cyberforensic methods for investigating
possible quantum cyberattacks on quantum communications and computation
networks;
\item Develop research into protecting automation softwares for quantum
gate definition in automated quantum communications protocols;
\item Develop solutions for protecting quantum networks against such types
of attacks;
\item Address the ways for remote deployment of the type of malware addressed
in the present work, as well as the possibility of insider threat,
which may be the greatest threat in the foreseeable future.
\end{enumerate}
On the detection side, the possibility of sending a prearranged test
message as a prefix key could be used to check for hacking of quantum
communications, making it possible a faster identification of attacked
nodes if measurements are periodically taken on specific nodes to
check for anomalies. 

On the other hand, since, in the examples reviewed, we are dealing
with a specific type of cyberattack in the form of malware that introduces
software-level changes to quantum machine code automation and hardware
implementation automation, with the objective of disrupting the communications
protocol, by introducing a modified sequence of unitary operations
depending on the messages that are being sent, so that the communications
protocol is disrupted, in this way disrupting quantum communications
and possible networked quantum computations, a counter for such an
attack may involve the need to develop antivirus solutions aimed at
checking this automation software for changes in the automatically
coded circuits to be implemented by the quantum hardware, under the
automated communications protocol.

As long as the antivirus is not itself hacked, this may provide for
an immediate way to identify a change in automated gate sequence software
and protect against the types of attacks addressed in the present
work.

In the near future, remote cloud-based access to quantum computers
can already be disrupted by such attacks as the ones addressed in
the present work, aimed at rising the error level of quantum computations
ran on these remote accessed systems, in this case, the threat is
of change in coded translation from high level programming languages
such as Python\textquoteright s Qiskit, for instance, into quantum
machine language instructions remotely sent to the hardware, if this
link is changed so that, for instance, a gate\textquoteright s specification
is randomly changed at specific intervals (for instance, an \emph{X}
gate to a unit gate), then one could raise the error rate of these
remotely accessed quantum computations, here the insider threat, aimed
at hindering a company or other organization running the quantum hardware
access is the greatest and near-future more feasible threat.

While the technical side of quantum cybersecurity may be concerned
with the patterns and different types of attacks that can disrupt
quantum communications and quantum networked computation, such a focus
is incomplete if one does not consider the way in which these attacks
can be operationally implemented by human beings, which leads us to
a direct intersection with Intelligence Studies.

In our case, since the attack target is software, the hacker needs
to conduct intelligence activities around the software systems used
for quantum gate automation, and to find ways in which to deploy the
malware, this intelligence gathering process and malware deployment
will depend upon the organization, but, given the foreseeable nature
of short to mid-term quantum communications, HUMINT activities will
most probably be involved, heightening the insider threat. 

Indeed, besides the threat from disgrunted employees that may deploy
the malware, hacker teams, whether state-sponsored or not, may compromise
people working for a target organization to gather intelligence and
deploy malware. Another possibility is to infiltrate an organization
using a covert operative to gather the intelligence on the gate specification
and automation software and to find ways in which to break that organization\textquoteright s
security protocols in order to deploy such a malware.

As quantum communications and quantum networked computation become
increasingly feasible, with the possibility of developping hybrid
classical/quantum networks and even fully quantum intranets that may
come to play a critical role in corporate, academia, defense, security
and intelligence organizations. These systems will become strategically
valued targets, especially for competitor states that are involved
in a race for quantum supremacy, in this regard, threat coming from
HUMINT activities targeting critical systems involved in quantum communications
and in networked quantum computation will become a matter of concern
from a National Security standpoint, especially taking into account
the types of organizations that may be involved in the use of advanced
quantum technologies. Future systematic studies are needed linking
quantum cyberattacks and quantum malware patterns with Intelligence
Studies in finding ways to deploy these attacks and how to counter
them.

\section*{References}
\begin{description}
\item [{Abellan,}] C. and Pruneri, V. (2018). \textquotedblleft The future
of cybersecurity is quantum\textquotedblright . IEEE Spectrum, 55(7),
30-35. DOI: 10.1109/MSPEC.2018.8389185.
\item [{Bennett}] C.H. and Wiesner S.J. (1992). \textquotedblleft Communication
via one- and two-particle operators on Einstein-Podolsky-Rosen states\textquotedblright .
Phys. Rev. Lett. 69, 2881. DOI: 10.1103/PhysRevLett.69.2881.
\item [{Cross,}] A.W., Bishop L.S., Smolin J.A., Gambetta J.M. (2017).
Open Quantum Assembly Language. arXiv:1707.03429 {[}quant-ph{]}. Retrieved
from: https://arxiv.org/pdf/1707.03429.pdf.
\item [{Cross}] A. (2018). \textquotedblleft The IBM Q experience and QISKit
open-source quantum computing software\textquotedblright . Bull. Am.
Phys. Soc. 63.
\item [{Gerhardt}] I., Liu Q., Lamas-Linares A., Skaar J., Kurtsiefer C.,
Makarov V. (2011). \textquotedblleft Full-field implementation of
a perfect eavesdropper on a quantum cryptography system\textquotedblright .
Nature Communications, 2:349. DOI: 10.1038/ncomms1348.
\item [{Gonçalves,}] C.P. (2017). \textquotedblleft Quantum Neural Machine
Learning: Backpropagation and Dynamics\textquotedblright . NeuroQuantology,
15, (1), 22-41.
\item [{Gonçalves,}] C.P. (2019). \textquotedblleft Quantum Neural Machine
Learning - Theory and Experiments\textquotedblright . In Aceves-Fernandez,
M.A. (ed.), Machine Learning in Medicine and Biology, IntechOpen,
DOI: 10.5772/intechopen.84149.
\item [{Gompert,}] D.C. and Libicki M. (2021). \textquotedblleft Towards
a Quantum Internet: Post-pandemic Cyber Security in a Post-digital
World\textquotedblright . Survival -- Global Politics and Strategy,
63(1), 113-124. DOI: 10.1080/00396338.2021.1881257.
\item [{Hugues-Salas}] E., Ntavou F., Ou Y., Kennard J. E., White C., Gkounis
D., Nikolovgenis K., Kanellos G., Erven C., Lord A., Nejabati R.,
Simeonidou D. (2018). \textquotedblleft Experimental Demonstration
of DDoS Mitigation over a Quantum Key Distribution (QKD) Network Using
Software Defined Networking (SDN)\textquotedblright . Optical Fiber
Communication Conference, OSA Technical Digest (Optical Society of
America), paper M2A.6, retrieved from: https://www.osapublishing.org/viewmedia.cfm?uri=OFC-2018-M2A.6\&seq=0.
\item [{Jogenfors}] J., Elhassan A.M., Ahrens J., Bourennane M., Larss
J.-Å. (2015). \textquotedblleft Hacking the Bell test using classical
light in energy-time entanglement--based quantum key distribution\textquotedblright .
Science Advances (1) 11: e1500793. DOI: 10.1126/sciadv.1500793. 
\item [{Larsson}] J.-Å. (2002). \textquotedblleft A practical Trojan Horse
for Bell-inequality-based quantum cryptography\textquotedblright .
Quantum Inf. Comput. 2, 434-442.
\item [{Lydersen}] L., Wiechers C., Wittmann C., Elser D., Skaar J., Makarov
V. (2010). \textquotedblleft Hacking commercial quantum cryptography
systems by tailored bright illumination\textquotedblright . Nat. Photon.
4, 686-689.
\item [{Makarov}] V. and Hjelme D. (2005). \textquotedblleft Faked states
attack on quantum cryptosystems\textquotedblright . Journal of Modern
Optics, 52(5), 691-705.
\item [{Mailloux,}] L. O., Hodson, D. D., Grimaila, M. R., McLaughlin,
C. V, Baumgartner, G. B. (2016a). \textquotedblleft Quantum Key Distribution:
Boon or Bust\textquotedblright . Journal of Cyber Security \& Information
Systems, 4(2), 18-25.
\item [{Mailloux,}] L.O., Lewis II C.D., Riggs C., Grimaila M.R. (2016b).
\textquotedblleft Post-Quantum Cryptography: What Advancements in
Quantum Computing Mean for IT Professionals\textquotedblright . IT
Professional, 18(5), 42-47.
\item [{Satoh}] T., Nagayama S., Suzuki S., Matsuo T., Van Meter R. (2020).
\textquotedblleft Attacking the Quantum Internet\textquotedblright .
arXiv:2005.04617 {[}quant-ph{]}, retrieved from: https://arxiv.org/pdf/2005.04617.pdf.
\item [{Satoh}] T., Nagayama S., Oka T. and Van Meter R. (2018). \textquotedblleft The
network impact of highjacking a quantum repeater\textquotedblright .
Quantum Science and Technology 3(3): 034008.
\item [{Schartner}] P. and Rass S. (2010). \textquotedblleft Quantum key
distribution and Denial-of-Service: Using strengthened classical cryptography
as a fallback option\textquotedblright . International Cpmputer Symposium
(ICS), Taiwan.
\item [{Wu}] L.-A. and Lidar D. (2006). \textquotedblleft Quantum Malware\textquotedblright .
Quantum Information Processing 5: 69--81. DOI: 10.1007/s11128-006-0014-5.
\item [{Zygelman}] B. (2018). \textquotedblleft A First Introduction to
Quantum Computing and Information\textquotedblright . Switzerland,
Springer. 
\end{description}

\end{document}